\documentclass[%
 reprint,
superscriptaddress,
 amsmath,amssymb,
 aps,
 prd,
]{revtex4-2}

\usepackage{graphicx}
\usepackage{dcolumn}
\usepackage{bm}
\usepackage{hyperref}

\usepackage{siunitx}
\usepackage{physics}
\usepackage{booktabs}
\usepackage{cleveref}
\usepackage{xcolor}
\usepackage{xspace}
\usepackage{aas_macros}
\usepackage{orcidlink}
\usepackage{makecell}

\graphicspath{{figs/}{./}}

\newcommand*{\mr}[1]{{\mathrm{#1}}}  
\newcommand*{\DM}{\ensuremath{\mr{DM}}\xspace}

\newcommand*{\figm}{\ensuremath{f_\mr{d}}\xspace}
\newcommand*{\Ob}{\ensuremath{\Omega_\mr{b}}\xspace}
\newcommand*{\Obf}{\ensuremath{\Omega_\mr{b}h^2\figm}\xspace}
\newcommand*{\DL}{\ensuremath{D\!_L}\xspace}
\newcommand*{\DMMW}{\ensuremath{\DM_\mr{MW}}\xspace}

\newcommand*{\DMc}{\ensuremath{\DM_\mr{cosm}}\xspace}
\newcommand*{\DMh}{\ensuremath{\DM_\mr{host}}\xspace}
\newcommand*{\DMe}{\ensuremath{\DM_\mr{exc}}\xspace}
\newcommand*{\sten}{\ensuremath{\sigma_\mr{host}}\xspace}
\newcommand*{\pJ}{\ensuremath{p_\mr J(H_0)}\xspace}

\newcommand*{\LCDM}{\ensuremath{\Lambda}CDM\xspace}

\newcommand*{\GWone}{GW170817\xspace}

\DeclareSIUnit \pc {pc}
\DeclareSIUnit \Mpc {\mega\pc}
\DeclareSIUnit \erg {erg}
\DeclareSIUnit \Jy {Jy}
\DeclareSIUnit \dmu {\per\cm\cubed\pc}
\DeclareSIUnit \yr {yr}
\DeclareSIUnit \Msol {\ensuremath{\mr{M}_\odot}}
\DeclareSIUnit \gauss {G}

\begin{document}

\preprint{APS/123-QED}

\title{Breaking the Baryon Density--Hubble Constant Degeneracy in Fast Radio Burst Applications with Associated Gravitational Waves}

\author{Joscha N.\ Jahns-Schindler\orcidlink{0000-0003-4193-6158}}
 \email{jjahnsschindler@swin.edu.au}
 \affiliation{%
 Max-Planck-Institut für Radioastronomie,
 Auf dem Hügel 69, 53121 Bonn, Germany
}%
 \affiliation{Centre for Astrophysics and Supercomputing, Swinburne University of Technology, Hawthorn, VIC 3122, Australia}
 \affiliation{ARC Centre of Excellence for Gravitational Wave Discovery (OzGrav), Hawthorn, VIC 3122, Australia}
\author{Laura G. Spitler\orcidlink{0000-0002-3775-8291}}%
\affiliation{%
 Max-Planck-Institut für Radioastronomie,
 Auf dem Hügel 69, 53121 Bonn, Germany
}%

\date{\today}

\begin{abstract}
Fast Radio Bursts (FRBs) are a unique probe of the cosmos, owing to dispersion caused by free electrons in the intergalactic medium (IGM).
Two of the main quantities of interest are degenerate: the density of matter $\Ob\figm$ outside of galaxies and the Hubble constant $H_0$.
Here, we present a new possibility of breaking the degeneracy without invoking early Universe priors on $\Ob$.
Assuming some FRBs originate in compact object mergers, the combination of dispersion and luminosity distance from the gravitational wave (GW) can be used to measure $\Obf$ (where $h$ is the dimensionless Hubble constant).
We show that this measurement can be combined with the abundant FRBs that have a redshift measurement.
This combination breaks the degeneracy with the Hubble constant.
We develop a Bayesian framework and forecast that third-generation GW detectors are required to obtain meaningful constraints.
We forecast that one year of Einstein Telescope operations can constrain $H_0$ to $\pm\SI{6}{\km\per\s\per\Mpc}$ and $\Obf$ to $^{+0.0015}_{-0.0016}$ (68$\,\%$ credible interval).
The method can also be used with luminosity distances obtained through other means than GWs.
\end{abstract}


\maketitle

\section{Introduction}
\label{sec:intro}

The disagreement between measurements of the local expansion rate of the Universe and the Hubble constant inferred from early Universe probes is known as the Hubble tension and poses one of the biggest problems to the standard cosmological model \citep{Riess2022,DESI2024}\citep[see][for recent reviews]{Abdalla2022,Hu2023}.
Because systematics are the leading order uncertainties, independent probes are essential to judge the significance of the disagreement.
One such proposed probe are Fast Radio Bursts (FRBs).
The dispersion measure (\DM) of FRBs is the integrated column density of electrons along their path.
As such, it is a proxy for distance that can be compared to the FRBs' host galaxy redshift ($z$).
The relation depends on the geometry of the Universe and hence can be used to infer cosmological parameters.
However, the $\DM$-$z$ relation depends on the combination $\Ob \figm H_0$, with $\Omega_\mr b$ the density parameter for baryons and $\figm$ the fraction of diffuse baryons in the IGM and in galaxy halos.
\citet{Macquart2020} constrain $\Ob h$ using a fixed $\figm$ from taking the remainder of a census of baryons in galaxies, where we  write the $h=H_0/\SI{100}{\km\per\s\per\Mpc}$.
\citet{James2022b} break the degeneracy further by assuming $\Ob h^2=0.02242$ from \citet{Planck2020}.
In this paper, we propose a new method to measure $\Ob h^2\figm$ using a subset of FRBs, proposed to be associated with Gravitational Waves (GWs).

The merger of two neutron stars (NSs) or of a NS with a black hole (BH) could potentially produce a GW and an FRB.
Although, NSs are now the favoured progenitor for the majority of the FRB population \citep{Bochenek2020, CHIME2020a}, mergers could still cause a fraction of FRBs.
The volumetric BNS merger rate of 10--\SI{1700}{\per\giga\pc\cubed\per\yr} \cite{Abbott2023} 
allows them to account for $\sim$\SI{2}{\%} of the observed FRBs, which have a rate of $7.3^{+8.8}_{-3.8}\times10^4\,\si{bursts}\,\si{\per\giga\pc\cubed\per\yr}$ above an energy of \SI{e39}{erg} (not corrected for beaming) \citep{Shin2023}.
The accounted fraction could be higher if the coalescence or the merger product produce more than one FRB.
Theoretically, FRBs have been suggested to arise from NS mergers in over four different ways \citep[see][for a comprehensive review]{Zhang2024}.
The interactions of magnetic fields before the merger might result in an FRB \citep{Lipunov1996, Metzger2016}.
A jet that is launched during the merger could interact with the surrounding medium and produce an FRB \citep{Usov2000}.
The NS resulting from the merger might emit FRBs in any way that has been suggested for NSs.
Finally, the newly formed NS can collapse into a BH if its mass exceeds a certain threshold.
When this happens, its whole magnetic field snaps and recombines, which might result in an FRB \citep{Falcke2014}.
Additionally, NS-BH mergers have been proposed as an FRB source through the interaction of the BH with the NS magnetic field \citep{Mingarelli2015,Bhattacharyya2017,Abramowicz2017}.

Several observational studies have been conducted to search for associated GW-FRB events.
\citet{Rowlinson2019} put constraints on FRB counterparts of GRB~150424A with triggered observations of the Murchison Widefield Array \citep[MWA][]{Tingay2013}.
\citet{Moroianu2022} reported a tentative association between GW190425 and FRB~190425A.
\citet{Panther2022} could also identify the most probable host of FRB~190425A to be the galaxy UGC10667 with \SI{80}{\percent} confidence.
However, \citet{Bhardwaj2023} re-estimate the GW parameters under the assumption that UGC10667 is the host.
They argue that an FRB could not pass through the ejecta of a merger after only 2.5 hours without noticeable attenuation.
They further infer a viewing angle of $>\ang{30}$ on the system, disfavouring the FRB emission in our direction.
Additionally, \citet{Radice2024} argue that at the location and distance of FRB~190425A, a kilonova would have been observed.
Most recently, \citet{Rowlinson2023} reported a tentative association of GRB~201006A with an FRB found with the Low Frequency Array \citep[LOFAR,][]{Haarlem2013}.
However in summary, no sufficiently significant association has yet been made that would allow establishing a definitive connection between GWs and FRBs.

Like classical probes, BNS merger as standard sirens can be used to measure $H_0$.
Here, the luminosity distance obtained from the chirp is compared to the redshift of the host galaxy.
This way, \citet{Abbott2017b} obtained $H_0=70^{+12}_{-8}\,\si{\km\per\s\per\Mpc}$ from \GWone.
The campaign around \GWone{} benefited from the fortunate circumstance that the event happened at a relatively close distance of only \SI{40}{\mega\pc} \citep[$z=0.01$][]{Abbott2017}.
Gradual improvements of GW experiments and eventually the third-generation detectors, Einstein Telescope and Cosmic Explorer \citep{Punturo2010,Evans2021}, will detect GWs to greater distances.
This will pose challenges for multi frequency follow-up campaigns similar to those conducted to find \GWone.
It comes as a convenient advantage that most telescopes used for FRB searches come with a large field of view.
ASKAP for example has a $\SI{30}{\square\deg}$ field of view and could almost cover the 31-$\si{\square\deg}$ initial LIGO-Virgo localisation region of \GWone in a single pointing.
Should BNS merger emit FRBs, these could lead future multi-messenger follow-up.

\citet{Wei2018} have previously developed a method \citep[largely based on][]{Gao2014} to use luminosity distances, \DL, from GW--FRB associations together with \DM{}s and redshifts to measure $\Omega_\mr{m}$ and $w$ in a $w$CDM cosmology.
\citet{Li2019a} proposed to employ the ratio of $\DL/\DM$ \citep{Gao2014} to measure $\figm$ using a minimum $\chi^2$-statistic.

Motivated by the tentative associations with unclear hosts, and since there will always be FRBs whose redshift cannot be obtained \citep{Marnoch2023,Jahns2023a}, we developed a new method to use GW--FRB associations that does not require a redshift.
We derive the full solution for the DM--\DL relation by inverting $\DL(z)$.
We only use the \DM and \DL of a subpopulation of FRBs and combine it with normal FRBs with measured redshifts.
This combination breaks the degeneracy in the \DM-$z$ relation and yields $H_0$ and $\Obf$ separately.

\section{Methods}
\label{sec:math}

The method we propose is based on the measurement of the \DM from the FRB and the luminosity distance, \DL, derived from the GW.
Both quantities are imprecise distance measurements.
The leading uncertainty in \DL comes from the degeneracy with the inclination angle of the merger system.
The reason is that the signal from a farther, face-on merger system can have a very similar shape as a closer but more edge-on system.
The measured dispersion is a composite of contributions from the host galaxy \DMh, the diffuse cosmic plasma in the IGM and the circumgalactic medium of intervening galaxies \DMc, and the Milky Way \DMMW, including its halo \citep{Deng2014}:
\begin{equation}
\DM = \DMMW + \DMc + \frac{\DMh}{1+z}\,. \label{eq:DMsum}
\end{equation}
The distance of the FRB can be estimated using the sight line-averaged $\langle\DMc\rangle(z)$.
The leading uncertainties in \DM stem from sight line-to-sight line variations in the baryon density (e.g.\ how many filaments are intersected) and from the uncertain \DMh.
We will marginalise over the probability distribution functions of \DL and \DMc in a Bayesian framework. But first, we derive the sight line-averaged $\langle\DMc\rangle(\DL)$.

\subsection{The \DM--\DL relation}
We assume a flat $\Lambda$-cold-dark-matter (\LCDM) universe where the redshift-dependent Hubble parameter is given by the Friedmann equation \citep[e.g.][]{Peacock1999} in the form
\begin{equation}
H(z)=H_0\sqrt{\Omega_\mr m(1+z)^3+\Omega_\Lambda}\,,  
\end{equation}
where $H_0$ is the Hubble constant, $\Omega_\mr m=\frac{\rho_\mr m}{\rho_0}$ the present-day matter density in terms of the critical density $\rho_0=\frac{3H_0^2}{8\pi G}$ with the gravitational constant $G$, and $\Omega_\Lambda=\frac{\rho_\Lambda}{\rho_0}$ the dark energy density, which can be expressed in terms of the cosmological constant $\Lambda$ as $\rho_\Lambda=\Lambda c^2/(8\pi G)$.
The luminosity distance for a source at redshift $z$ follows as
\begin{equation}
\DL(z) 
=(1+z)\frac{c}{H_0}\int_0^z \frac{1}{\sqrt{\Omega_\mr m(1+z)^3+\Omega_\Lambda}}\dd{z}\,.
\label{eq:DL}
\end{equation}

The average of \DMc over many sight lines is given as \citep{Ioka2003,Inoue2004,Deng2014}
\begin{equation}
\langle\DMc\rangle(z) = \frac{3c}{8\pi Gm_\mr{p}} \Omega_\mr b H_0  \int_0^z\frac{(1+z)\,\figm(z)\,\chi(z)}{\sqrt{\Omega_\mr m(1+z)^3+\Omega_\Lambda}}\dd{z}\,, \label{eq:Macquart2}
\end{equation}
where $m_\mr{p}$ is the proton mass and $\chi(z)$ the ionisation fraction at a given redshift, which can be expressed in terms of the ionisation fractions of hydrogen and helium $\chi(z)=\frac{3}{4}\chi_\mr{H}(z)+\frac{1}{8}\chi_\mr{He}(z)$.
In this work, we will consider only low-$z$ FRBs, so we assume hydrogen and helium to be fully ionised, i.e., $\chi_\mr{H}(z)=\chi_\mr{He}(z)=1$, and we assume \figm is constant.
Eq.~(\ref{eq:Macquart2}) then becomes
\begin{eqnarray}
\langle\DMc\rangle(z) = &&\frac{3c}{8\pi Gm_\mr{p}}\frac{7}{8} \Omega_\mr b \figm H_0 \nonumber\\
&&\times\int_0^z\frac{1+z}{\sqrt{\Omega_\mr m(1+z)^3+\Omega_\Lambda}}\dd{z}\,.\label{eq:Macquart3}
\end{eqnarray}
This famous \DM--$z$ relation depends on the product $\Omega_\mr b \figm H_0$, such that these quantities can not be inferred individually.
This degeneracy can only be broken with additional constraints, e.g., from measurements of the cosmic microwave background.

If only \DM and \DL are available while $z$ is not directly measurable, we can invert $\DL(z)$ to obtain
$\langle\DMc\rangle(\DL)$.
The inversion of $\DL(z)$ can only be done numerically, but this does not hinder us from using the relationship.
We will refer to the inverted function as $z(\DL)$ and write
\begin{equation}
    \langle\DMc\rangle(\DL)=\langle\DMc\rangle(z(\DL))\,.
\end{equation}
The Taylor approximation can be done analytically; the first-order term is
\begin{equation}
\langle\DMc\rangle(\DL) 
\approx  \frac{3}{8\pi Gm_\mr{p}}\frac{7}{8} \Omega_\mr b \figm H_0^2 \,\DL\,.
\end{equation}
Interestingly, this equation is proportional to $\Omega_\mr b \figm H_0^2$, while the classic \DM--$z$ relation has a proportionality to $\Omega_\mr b \figm H_0$.
Combining the \DM--\DL relation with the \DM--$z$ relation, i.e., GW-FRB associations with independent FRBs that have a measured redshift, we can break the degeneracy between  $\Omega_\mr b \figm$ and $H_0$.

\begin{figure}
	\centering
	\includegraphics[width=\hsize]{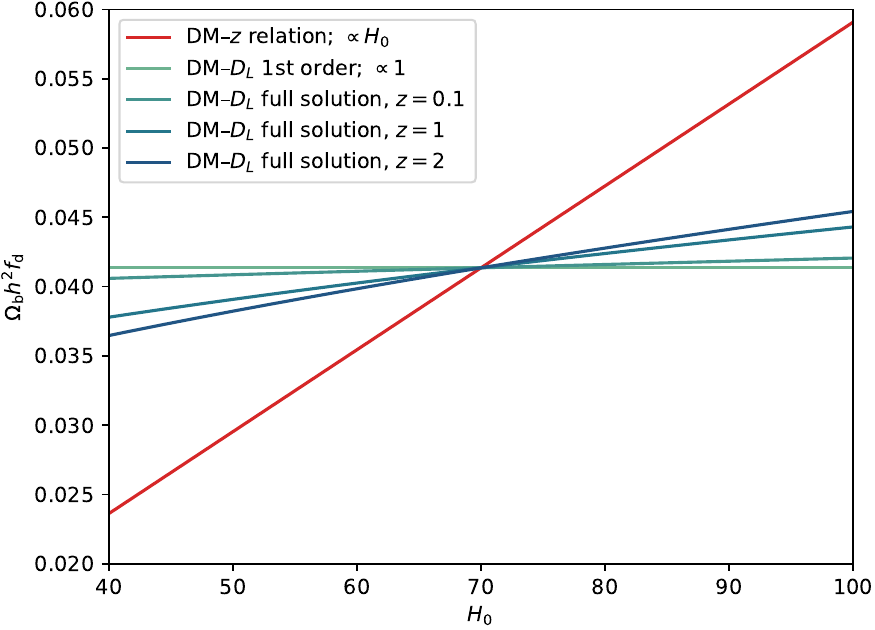}
	\caption[Curves of parameter degeneracy.]{Curves along which parameters are degenerate in different methods. 
    Measuring both relations breaks the degeneracy by allowing only the intersection of the two curves.}
	\label{fig:Obf_H0}
\end{figure}

Fig.~\ref{fig:Obf_H0} illustrates the curves along which perfectly measured relations would still allow parameters to be free. If the \DM--$z$ and \DM--\DL relations are both measured, only the intersection of the two relations is allowed and the degeneracy is broken.
The degeneracy can also be broken by using only the \DM--\DL relation with FRBs at different redshift.
The reason is the deviation of the full solution from the linear relation.
Due to large uncertainties in single measurements, this variation will only give useful constraints long after the combination with the \DM--$z$ relation.

\subsection{Bayesian framework}
\label{sec:bayes}

To utilise the derived relation and combine it with other FRB data, we use a Bayesian framework.
From Bayes rule \citep[see e.g.][for an introduction]{Jaynes2003}, the posterior probability distribution is proportional to the product of the likelihood and the prior probability:
\begin{alignat}{1}
p(&\Obf , H_0\mid \DM, \DL) \nonumber \\
&\propto p(\DM, \DL\mid\Obf, H_0) \, p(\Obf, H_0) \nonumber \\
&=\prod_i p(\DM_i, {\DL}_i\mid\Obf, H_0) \, p(\Obf, H_0)\,,\label{eq:bayes}
\end{alignat}
where $i$ goes over all FRBs.

In this exploratory work, we will restrict ourselves to the two free parameters \Obf and $H_0$, but additional parameters should be left free in further studies.
We assume that the Milky Way contribution can be subtracted from the measured \DM with sufficient precision, such that we can use the resulting excess $\DMe=\DM-\DMMW=\DMc+\DMh/(1+z)$.

We follow \citet{Macquart2020} in the likelihood definitions of \DMc and \DMh.
We describe \DMh by a log-normal PDF in the formulation of \citet{Jahns2023a} with median $\DM_0$ and the standard deviation $\sten$ of the \DM's base 10 logarithm, i.e.:
\begin{alignat}{1}
p_\mr{host}(\DMh & \mid\DM_0,\sten) = \frac{\log_{10}(e)}{\DMh\sten\sqrt{2\pi}} \label{eq:pDMh} \\
&\times \exp\left(-\frac{(\log_{10}\DMh-\log_{10}\DM_0)^2}{2\sten^2}\right)\,. \nonumber
\end{alignat}
We describe the PDF of \DMc for a given \DL in terms of $\Delta=\frac{\DMc}{\langle\DMc\rangle(z(\DL))}$ and with the fluctuation parameter $F$ as
\begin{alignat}{1}
p_\mr{cosm} (\DMc\mid & \DL,\Obf,H_0,F) = \nonumber\\
&\frac{A\Delta^{-3}}{\DMc}\exp\left(-\frac{(\Delta^{-3}-C_0)^2}{18\,\sigma^2}\right)\,,
\label{eq:pDMc}
\end{alignat}
where $\sigma=F/\sqrt{z(\DL)}$, $A$ is the normalisation constant and $C_0$ is fixed by the condition $\langle\Delta\rangle = 1$.  For the assumed flat universe, the normalisation can be calculated analytically and is given in the appendix.

Because $\DMh=(\DMe-\DMc)\times(1+z)$, the above equations still leave some freedom for either \DMc or \DMh.
We, therefore, use a flat prior for \DMc between 0 and \DMe and marginalise over it.
Additionally, \DL is not an exact measurement but described by a posterior distribution resulting from the GW analysis.
When simulating data, we assume that the uncertainty of \DL follows a normal distribution.
Finally, we incorporate the constraints on $\Ob H_0\figm$ from FRBs with redshifts (and without GWs) 
as a prior in the form 
$p_{H_0}(H_0\mid \Obf)$
and put a flat prior distribution for $\Obf$.
Collecting everything in Eq.~(\ref{eq:bayes}) and integrating over \DL and \DMe, yields
\begin{widetext}
\begin{alignat}{2}
p(&\DM_i, {\DL}_i\mid\Obf, H_0) \, p(\Ob && h^2\figm, H_0) = \int \dd{\DL} \int_0^{\DMe} \dd{\DMc} \, p(\DMc, \DL\mid\Obf, H_0) \, p(\Obf, H_0)\nonumber\\
& = \int \dd{\DL} \int_0^{\DMe} \dd{\DMc} \, && p_\mr{cosm}(\DMc\mid \DL,\Obf,H_0,F) \, p_\mr{host}(\DMe-\DMc\mid\DM_0,\sten) \nonumber \\
& &&\times p_{H_0}(H_0\mid \Obf) \,p(\DL)\,p(\Obf) 
\end{alignat}
\end{widetext}
where $F$ is fixed but can be marginalised over in future work, and $\DM_0$, and \sten are fixed at first but will be left free later.

\section{Forecasts}
\label{sec:simulations}

Now, we will simulate a population of GW-FRB events to forecast the potential of this method.
We will explore the posterior distribution using Markov chain Monte Carlo (MCMC) simulations using the \textsc{emcee} package \citep{ForemanMackey2013}.

We make three forecasts.
One resembles future detections of current GW detectors and one the next generation GW detectors assuming the specifications of the proposed Einstein Telescope.
The third run is similar to the second but treats the parameters of the \DMh distribution as unknowns.
We combine the first forecast with the current constraints from FRBs and the second and third with the projected constraints of future FRBs. Detections with current GW detectors will mainly stem from observing run O5 of the international gravitational-wave network, which is planned for 2027-2030\footnote{\url{https://observing.docs.ligo.org/plan/#timeline}}.
The Einstein Telescope is planned to start operations in 2035 \citep{Maggiore2020}.

\subsection{Current generation GW detectors}
\label{sec:current}

We simulate 10 FRBs with GW counterparts.
The true number is very uncertain.
\citet{Kiendrebeogo2023} predict the number of BNS mergers in O5 to be $86^{+171}_{-59}$ or $180^{+220}_{-100}\,\mr{yr}^{-1}$ (uncertainties are the 5\% and 95\% quantiles), for two different population models.
Additionally, it is uncertain how many of them could emit a detectable FRB.

For simplicity, we place all events at a fixed redshift $z=0.1$.
This is conservative in comparison to the forecasted \citep[table 4 in][]{Kiendrebeogo2023} median luminosity distances of $619^{+15}_{-19}$ and $738^{+30}_{-25}$\,\si{\Mpc} ($z=0.127$ and 0.150, respectively, uncertainties are again the 5\% and 95\% quantiles), because a higher median redshift means a higher cosmological signal relative to the uncertain contributions.

We assume a 1-$\sigma$ Gaussian \DL uncertainty of \SI{40}{\percent}, which is the median of presently detected GW events\footnote{Using the data from \url{https://gwosc.org}.}.
With this uncertainty, we draw the \DL mock measurements around the value corresponding to $z=0.1$.
In the inference, we consider it in the form of a normal likelihood for \DL.
To draw \DM, we use Equations~(\ref{eq:pDMh}), (\ref{eq:pDMc}), (\ref{eq:DMsum}).
We summarise all parameters in Tables~\ref{tab:GWparams} and \ref{tab:GWparams2}.

\begin{table}
	\centering
	\caption{Parameters used for our simulations and prior distributions used in the inference. Brackets denote the limits of flat distributions, $\mathcal{N}(\mu,\sigma)$ denotes the normal distribution.}
	\begin{tabular}{lr}
		\toprule
		Parameter	&	Simulation value	\\
		\midrule
		$F$	&	0.32	\\
		$\sten$	&	0.57	\\
		Cosmology	&	Flat \LCDM	\\
		$H_0$	&	\SI{73}{\km\per\s\per\Mpc}	\\
		$\Omega_\mr{m}$	&	0.3	\\
		\midrule
		 &	Prior\\
		\midrule
		$H_0$	&	$[10, 150]\,\si{\km\per\s\per\Mpc}$	\\
		$\Obf$	&	$[0, 1]$	\\
		\DMc	&	$[0, \DMe]$	\\
		\DL	&	$\mathcal{N}(\DL(z), \sigma_\DL)$	\\
		\bottomrule
	\end{tabular}
	\label{tab:GWparams}
\end{table}

\begin{table*}
	\centering
	\caption{Parameters and distributions that are different between the different runs.}
	\label{tab:GWparams2}
 	\begin{tabular}{lrrr}
		\toprule
        Parameter	&	LIGO+Virgo+KAGRA & Einstein Telescope & \makecell{Einstein Telescope\\free host DM parameters} \\ 
        \midrule
        Number & 10 & 20, 100 & 20, 100\\ 
		$z$	&	0.1 & 	0.1, 0.2 & $\mathcal{N}^+(0.1, 0.1)$, $\mathcal{N}^+(0.2, 0.2)$\\
        $\sigma_\DL/\DL$ & 0.4 & 0.1, 0.2 & 0.1, 0.2\\
		$\DM_0$	&	\SI{170}{\dmu}	& \SI{170}{\dmu} & \SI{100}{\dmu}\\
        FRB-$z$ prior & \citet{James2022b} & \multicolumn{2}{r}{$\mathcal{N}(73\,\si{\km\per\s\per\Mpc}, 0.8\,\si{\km\per\s\per\Mpc})$}\\ 
		\bottomrule
        \multicolumn{4}{l}{\footnotesize{$^+$ $\mathcal{N}^+$ is the truncated normal distribution, truncated at 0 to give only positive values.}}
	\end{tabular}
\end{table*}

As a prior from the `classic' FRB-$z$ method, we use the current constraints on $\Ob \figm H_0$ (which might significantly improve in the next years).
Specifically, we use the constraints from mostly localised FRBs from \citet{James2022b}, as detailed in the appendix.

\subsection{Einstein Telescope}

To forecast the potential for the next generation of GW detectors, we adhere to the forecasts for the Einstein Telescope of \citet{Iacovelli2022}.
They predict about 20 out of their $\sim$10000 annual detections to have a relative uncertainty in the luminosity distance below $10\,\%$ and about 100 below $20\,\%$.
As conservative values, we pick 10 and $20\,\%$ and the estimated median redshifts of $0.1$, and $0.2$, respectively.

As the FRB-$z$ prior, we assume a normally distributed PDF for $H_0$.
\citet{James2022b} predict an $H_0$ uncertainty of about $2.45\,\si{\km\per\s\per\Mpc}$ for 100 FRBs.
We extrapolate this value to $\sim$1000 FRBs through $2.45/\sqrt{10}\approx0.8$.
The actual number of localised FRBs will likely be higher by the time of the Einstein Telescope, but systematics are going to start dominating the uncertainties.

\subsection{Leaving the host DM free}

So far, we have assumed that \DMh in the GW-FRB population has the same distribution as \DMh of the general FRB population, this is unlikely if the progenitors are different.
The FRBs coming from mergers could, e.g., come from an older stellar population, which would also imply a lower \DMh compared to an FRB population of young magnetars.
This requires a separate inference of \DMh for the GW-FRB population.

We model \DMh as a log-normal distribution with parameters $\DM_0$ and \sten and leave them as free parameters in our inference.
As input values, we take $\DM_0=\SI{100}{\dmu}$ and $\sten=0.57$.
Because of the dependency on $z$ in Eq.~\ref{eq:DMsum}, a realistic spread in $z$ would now help in constraining the \DMh parameters and give more realistic forecasts.
We therefore distribute the sources in $z$.
We choose a normal distribution requiring values to be positive, with 20 FRBs with $z\sim\mathcal{N}(0.1, 0.1)$ and 100 FRBs with $z\sim\mathcal{N}(0.2, 0.2)$.
This choice is motivated again by fig.~15 of \citet{Iacovelli2022}.

\subsection{Results}
\label{sec:gwres}

\begin{figure}
	\centering
	\includegraphics[width=\hsize]{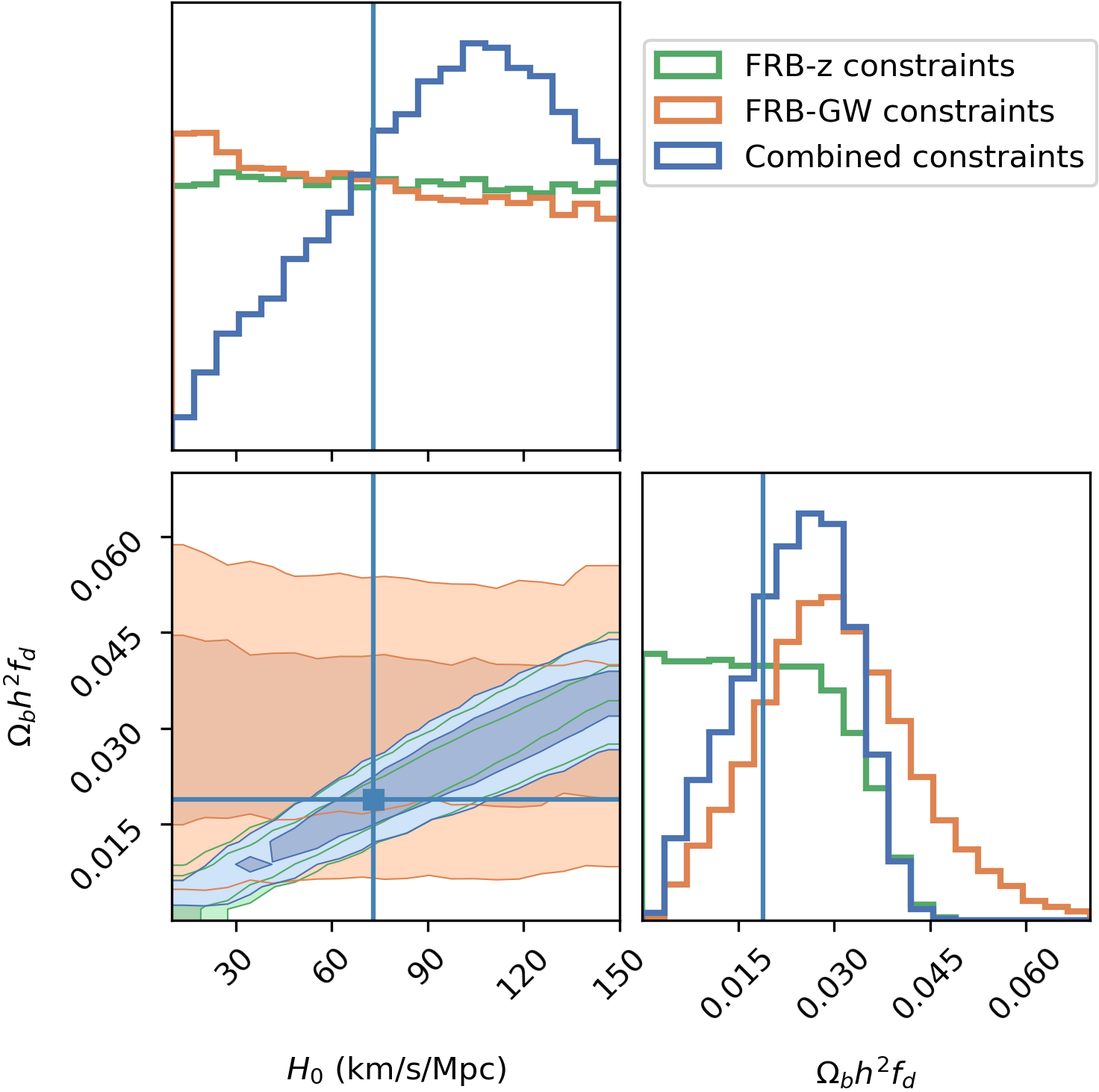}
	\caption[Corner plot with 10 simulated events.]{Posterior distribution when using 10 simulated FRB--GW events. The prior distribution $p_{H_0}$ derived from the \DM--$z$ relation \citep{James2022b} is also shown, as well as the posterior distribution without using this prior. Contours show the 68 and \SI{95}{\%} confidence regions. Panels above and right of the plot show the posteriors when marginalising over one of the variables. Blue lines mark the input values.}
	\label{fig:10FRBs}
\end{figure}
\begin{figure}
	\centering
	\includegraphics[width=\hsize]{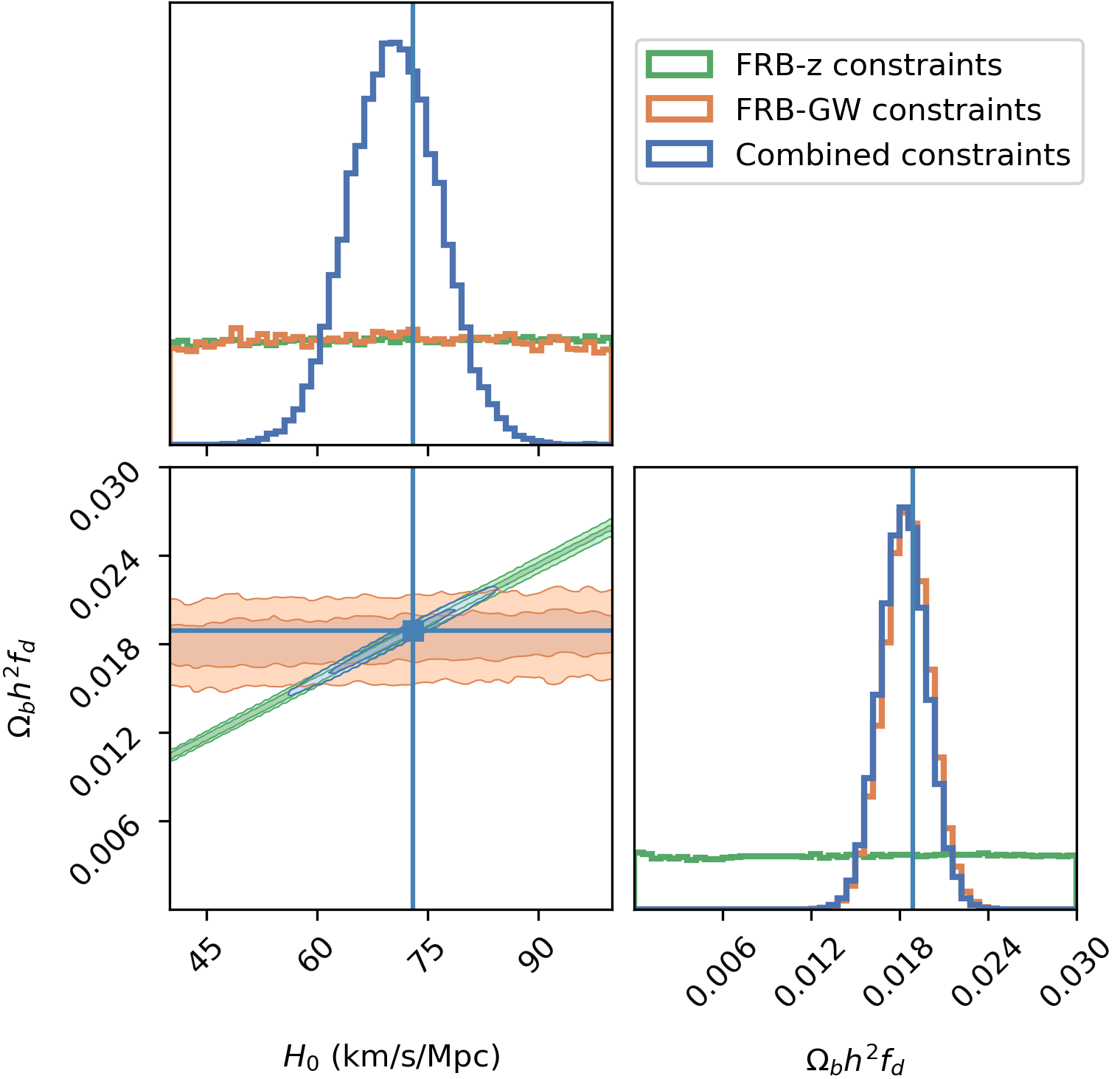}
	\caption[Corner plot with 120 simulated events.]{Same as Fig.~\ref{fig:10FRBs} but for 120 simulated FRB--GW events as specified in Tab.~\ref{tab:GWparams}.}
	\label{fig:100FRBs}
\end{figure}
\begin{figure}
	\centering
	\includegraphics[width=\hsize]{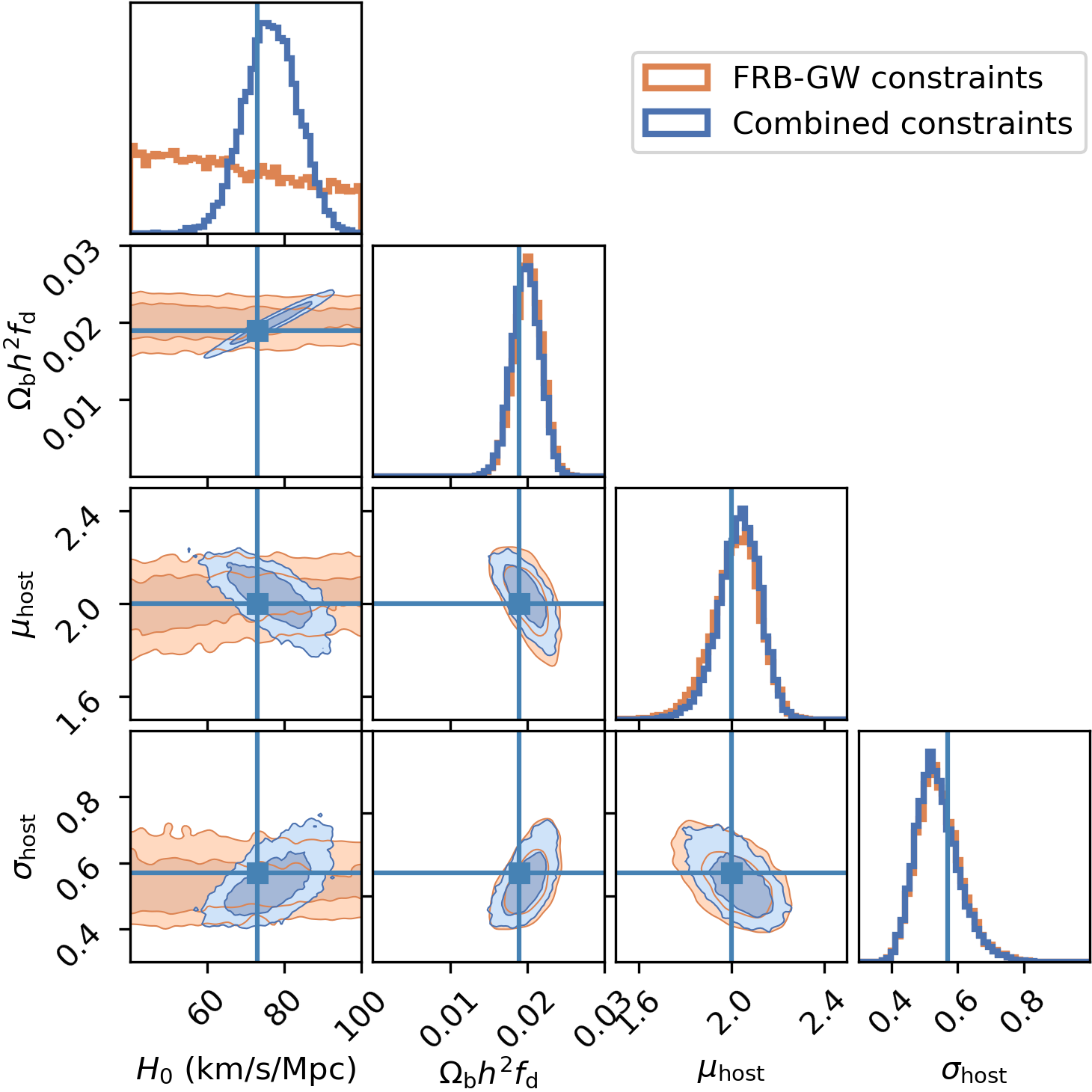}
	\caption[Corner plot with free host DM.]{Same as Fig.~\ref{fig:100FRBs} but treating the host parameters -- $\DM_0$ and \sten -- as unknown. The source redshifts are now distributed. The parameter $\mu_\mathrm{host}=\log_{10}\DM_0$.}
	\label{fig:free_host}
\end{figure}

We applied the Bayesian inference to the three simulated data sets.
The results are shown in Figures~\ref{fig:10FRBs}, \ref{fig:100FRBs}, and \ref{fig:free_host}.

The 10 simulated FRB--GW events at $z=0.1$ can not constrain the parameter space much beyond the prior distribution.
The reason is likely the large scatter in \DM and $\DL$.
Particularly at such low redshifts of only $z=0.1$, the median of \DMh is larger than the median of \DMc.

The 120 simulated FRB--GW events at $z=0.1$ and 0.2 show meaningful constraints.
The resulting parameters are $H_0=\SI{71(6)}{\km\per\s\per\Mpc}$ and $\Obf=1.83^{+0.15}_{-0.16} \times 10^{-2}$ with \SI{68}{\percent} credibility, an \SI{8.6}{\percent} and \SI{8.5}{\percent} relative uncertainty, respectively.

The case with free $\DM_0$ and \sten gives $\sigma_{H_0}=^{+7}_{-7}$ and $\sigma_{\Obf} =^{+0.19}_{-0.17}\times 10^{-2}$, a small increase with respect to the previous case where the parameters were fixed.

\section{Discussion}
\label{sec:discussion}
\subsection{Comparison to other probes}
The constraints on $H_0$ we forecast for 120 events are tighter than the current constraints from FRB \DM--$z$ measurements with CMB priors on $\Ob h^2$, which yield an uncertainty of $\sigma_{H_0}=^{+22}_{-12}\,\si{\km\per\s\per\Mpc}$ \citep[\SI{90}{\percent} credible;][]{James2022b}.
Compared to the assumed future constraints of $\sigma_{H_0}=\SI{0.8}{\km\per\s\per\Mpc}$, we are loosing a factor 7.5 in precision; the price we pay for independence from early Universe measurements.

Compared to current measurements, the predictions exceed some less precise, current probes like GWs \citep[$H_0=69^{+16}_{-8}\,\si{\km\per\s\per\Mpc}$ (\SI{68}{\%} credible),][]{Abbott2021}, but we would have to accumulate several years of data to reach the current precision of, e.g., Type Ia supernovae calibrated with Cepheids and parallax measurements \citep[\SI{73.04 \pm 1.04}{\km\per\s\per\Mpc},][]{Riess2022} or from Baryonic Acoustic Oscillations \citep[\SI{68.52 \pm 0.62}{\km\per\s\per\Mpc},][]{DESI2024}.
In the end, how well we can understand the systematics in FRB \DM{}s will be decisive.

\subsection{Reviewing assumptions}

The assumptions of the forecast were conservative in several ways.
The assumed distribution of \DMh from \citet{James2022b} given by $\DM_0=170^{+59}_{-47}\,\si{\dmu}$ (all uncertainties in this discussion are 68\% credible) and $\sigma_\mr{host}=0.57^{+0.13}_{-0.09}$ is high in comparison to the results of other studies.
\citet{Shin2023} for example find $\DM_0=84^{+69}_{-49}\,\si{\dmu}$ and $\sigma_\mr{host}=0.41^{+0.21}_{-0.20}$.
This would result in a much narrower PDF of \DMh and in turn much narrower posterior distributions.

Our assumption of vanishing \DMMW is not always warranted.
The halo contribution is known with an uncertainty of \SI{\sim10}{\dmu}, which is certainly much smaller than the uncertainty of \DMh and can be absorbed by it.
Ideally yet, it is modelled separately as, e.g., \citet{Hoffmann2024} do with $\DMMW\sim\mathcal{N}(\SI{50}{\dmu},\SI{15}{\dmu})$.
In particular at low galactic latitudes, model uncertainties in the MW ISM can be larger than the halo uncertainty and one problem is that the uncertainties themselves are not very well known \citep{Price2021}.
The effects from wrongly estimated \DMMW can be estimated using the approximation $\DMc\approx z\times\SI{1000}{\dmu}$.
Generally, an underestimated \DMMW will produce an estimate of \Obf and $H_0$ that is too high.
This potential bias is shared with all FRBs and is therefore one of the primary goals that will be addressed with close-by FRBs.

The uncertainty in each \DMc could be reduced by detailed optical follow-up.
By taking spectra of galaxies close to the sight line, one can attribute a part of the \DM to their halo and identify filaments or galaxy clusters along the sight line \citep{Simha2020, Lee2022, Khrykin2024}.

Although we derive it in the context of a GW-FRB association, the method presented in this paper only requires a luminosity distance measurement of any kind in combination with the \DM of an FRB.

\subsection{Preferred direction of FRB emission}

In this work, we implicitly assumed that the emission direction of the FRB is independent of the inclination of the merger system with respect to the observer.
If instead there is a preferred direction, the chances of simultaneous detection of FRBs and GWs are affected, and the \DL inference can be biased.
If FRBs are preferentially emitted towards the poles but isotropic emission is assumed, the inferred \DL is lower than the real value, resulting in \Obf and $H_0$ being biased high.

Some models for FRB emission from BNS mergers favour a direction, but often the direction is not explicitly discussed or not predicted.
In models where the FRB emerges from the jet \citep[][]{Goodman1986, Paczynski1986, Usov2000} or the GW itself \citep{Moortgat2003,McWilliams2011}, the direction of the rotational axis is preferred.
In models with pulsar-like emission \citep[e.g.][]{Totani2013}, the direction of the pulsar magnetic field is preferred, which is expected to align with the system's rotational axis during inspiral \citep{Cheng2014,Giacomazzo2015}; yet more complex magnetic fields might allow emission in other directions.
Ultimately, it is unclear which models can produce FRBs.

To find the correct prior, the beaming direction of the FRB with respect to the merger axis needs to be studied.
One possibility is to compare the FRB brightness with the brightness of other electro-magnetic counterparts, e.g., X-rays and gamma rays.

\subsection{Future detectors not considered here}

Apart from the Einstein Telescope considered here, additional 2.5 and third-generation GW detectors are being planned.
The Cosmic Explorer \citep{Evans2021} is planned to start operations in 2035.
\citet{Iacovelli2022} predict the BNS merger rates to increase from \numrange{910}{160000} per year for the Einstein Telescope alone to \numrange{3800}{650000} per year for the combination of Einstein Telescope and Cosmic Explorer.
The precision in the luminosity measurements will also increase and therefore the number of events with a $<$\SI{10}{\%} uncertainty from $\sim$20 to $\sim$ 2000.

In terms of FRB detections, the most suitable telescopes will be ones with a very wide field of view, like the Bustling Universe Radio Survey Telescope in Taiwan \citep[BURSTT,][]{Lin2022}.

\subsection{Mergers with redshift measurements}
In this work, we explicitly assumed events without measured redshifts.
If the redshift is measured, the \DL--$z$ relation could provide the $H_0$ measurement to directly break the degeneracy in the \DM--$z$ relation.
It would be straightforward to extend our Bayesian framework to include the redshift.

However, when assuming perfectly known redshifts, the likelihoods of \DM and \DL factorise and become independent.
This method would therefore be similar to combining the FRB data with some external $H_0$ measurements.
The joint analysis of \DL, \DM, and $z$ will be favourable, when additional cosmological parameters are considered or when the redshift uncertainty from peculiar velocities becomes relevant, which is often the case at low redshifts.

Previous works mentioned in the introduction have suggested other methods to use the three measurements of \DL, \DM and $z$.
\citet{Gao2014} propose $\DL(z)/\langle \DMc \rangle (z)$ as a probe of the scatter in $\DMc$ and the evolution in the scatter with redshift.
They argue that the division makes it less dependent on cosmological parameters because the integrals in Eq.~\ref{eq:DL} and Eq.~\ref{eq:Macquart3} almost cancel out.
\citet{Wei2018} suggested to instead use the product $\DL\times\langle\DMc\rangle(z)$, which is independent of $H_0$, to measure $\Omega_\mr{m}$ and $w$ in the $w$CDM model.
\citet{Li2019a} again use the ratio $\DL/\langle\DMc\rangle(z)$ but with the aim of measuring $\figm$, similar to this study.

At several points in theses papers, it is claimed that $z$ and $D_L$ can be used interchangeably and it is sometimes not clear which one is used in their mock observations. However, this is not true as calculating, e.g., $D_L$ from $z$ depends on cosmology so if only $z$ is measured, the \DM--$z$ relation needs to be used and vice versa.

Furthermore, all these methods take the product or ratios of the measurements, discarding information from the individual values.
Their advantages of being less dependent on $H_0$ or other cosmological parameters will also be there in a Bayesian framework.
Therefore, and because it is more flexible and clear, our Bayesian framework should generally be advantageous.

\section{Conclusion}
We developed and tested a new method to use FRBs with an associated luminosity distance measurement as a cosmological tool.
This luminosity distance could most likely come from a GW counterpart if FRBs arise during a BNS merger.
The \DM--\DL measurements allow inferring \Obf and $H_0$ when combined with independent \DM--$z$ measurements from localised FRBs.
The combination of measurements breaks the degeneracy between \Obf and $H_0$ that each measurement would have individually, without using external priors from cosmic microwave background or supernovae measurements.
We developed a Bayesian framework to use the \DM and \DL measurements and combine them with the existing \DM--$z$ measurements.
We find that third-generation GW detectors like the Einstein Telescope and the Cosmic Explorer are required to obtain meaningful constraints.
According to our simulations, 120 events could constrain $H_0$ and \Obf to $\SI{\pm6}{\km\per\s\per\Mpc}$ and $^{+0.15}_{-0.16} \times 10^{-2}$, respectively.
When assuming that the \DMh of FRBs that are associated with GWs is distributed differently than \DMh of the general FRB population, we obtain uncertainties of $^{+8}_{-8}$ and $^{+0.21}_{-0.20}\times 10^{-2}$ for $H_0$ and \Obf, respectively.

\vspace{1.5pt}
The simulation and inference code used in this article is available at \url{https://github.com/JoschaJ/DMDLinference}.

\begin{acknowledgments}
We want to thank Charles Walker for discussions in the early project phase.
We want to thank Norbert Wex for discussions about GWs and in particular \GWone at the start of the project.
Likewise, we want to thank Kristen Lackeos for discussions about future GW detectors.
We thank the referee for useful comments, especially for emphasising that a shared distribution of host DMs between the two populations is a bad assumption and for pointing out potential biases from preferred inclination angles.

JJ-S acknowledges support through Australian Research Council Discovery Project DP220102305.

LS is a Lise Meitner independent group leader and acknowledges funding from the Max Planck Society.

\end{acknowledgments}

\bibliography{frbs}

\appendix*

\section{Numerical simplification}
\label{app:math}
As we assumed a flat cosmology, we speed up our numerical calculations by using the analytic solutions of the two integrals in Eqs.~(\ref{eq:DL}) and (\ref{eq:Macquart3}).
These integrals can be calculated as
\begin{eqnarray}
	\int_0^z &&\frac{1}{\sqrt{\Omega_\mr m(z+z)^3+\Omega_\Lambda}}\dd{z} 
    = \frac{z+1}{\sqrt{\Omega_\Lambda}} \nonumber \\
    &&\times{_2F_1}\eval{\left(\frac{1}{3}, \frac{1}{2}; \frac{4}{3}; -\frac{\Omega_\mr m (1+z)^3}{\Omega_\Lambda}\right)}_0^z\quad\text{and}
\end{eqnarray}
\begin{eqnarray}
	\int_0^z && \frac{1+z}{\sqrt{\Omega_\mr m(z+z)^3+\Omega_\Lambda}}\dd{z} = \frac{(z+1)^2}{2\sqrt{\Omega_\Lambda}}\nonumber\\ &&\times\eval{{_2F_1}\left(\frac{1}{2}, \frac{2}{3}; \frac{5}{3}; -\frac{\Omega_\mr m (1+z)^3}{\Omega_\Lambda}\right)}_0^z\,,
\end{eqnarray}
where ${_2F_1}$ is the Gaussian hypergeometric function.

The normalisation for Eq.~\ref{eq:pDMc}
\begin{eqnarray}
	A && =  3(12\sigma)^{1/3}\Biggl[3\,\Gamma\left(\frac{1}{3}\right)\,\sigma{_1F_1}\left(\frac{1}{6}, \frac{1}{2}, -\frac{1}{2}\left(\frac{C_0}{{3 \sigma}}\right)^2\right) \nonumber \\
	&& + \sqrt{2}\,\Gamma\left(\frac{5}{6}\right)\, C_0\, {_1F_1}\left(\frac{2}{3}, \frac{3}{2}, -\frac{1}{2}\left(\frac{C_0}{{3 \sigma}}\right)^2\right)\Biggr]^{-1}\,,
\end{eqnarray}
where $\Gamma$ is the gamma function, and ${_1F_1}$ is the generalised hypergeometric function.

\section{Obtaining the prior for \Obf}
Here, we give the details on how we obtained the prior from current FRB-$z$ constraints.
Since \citet{James2022b} used priors on \Obf to constrain $H_0$, but we want to leave it as a free parameter, we have to exclude their prior again.
We read the PDF of $H_0$ from their fig.~6 using the \textsc{WebPlotDigitizer} \citep{Rohatgi2022}, as the data are not published.
\citet{James2022b} fix $\Ob h^2=0.02242$, using big bang nucleosynthesis theory and cosmic microwave background measurements \citep{Planck2020}.
They further use $\figm(z=0) =0.844$ with an evolving $\figm(z)$.
As we have no possibility to account for the evolution, we have to assume it did not have a big influence on their result, which is reasonable given the mostly low-$z$ FRBs.
Hence, we fix $\figm=0.844$ and obtain a constant $C$ that allows us to reverse their assumptions:
\begin{equation}
C=\Ob h^2 \figm = 0.02242\times 0.844\,.
\end{equation}
We then transform the PDF reported for $H_0$, which we label as $\pJ$, to account for the change of variables
\begin{equation}
p_{H_0}(H_0\mid \Obf) = \frac{C}{\Obf}\, p_\mr{J}\left(\frac{C}{\Obf}H_0\right)\,.
\end{equation}

\end{document}